\documentclass[aps,prc]{revtex4}
\usepackage{amssymb,amsmath,epsfig, psfrag, ulem, color} 

\newcommand{\rf}[1]{(\ref{#1})}

\newcommand{\ee}{\end{equation}}
\newcommand{\be}{\begin{equation}}
\newcommand{\bea}{\begin{eqnarray}}
\newcommand{\eea}{\end{eqnarray}}
\newcommand{\bean}{\begin{eqnarray*}}
\newcommand{\eean}{\end{eqnarray*}}

\newcommand{\gapproxeq}{\lower
.7ex\hbox{$\;\stackrel{\textstyle >}{\sim}\;$}}
\newcommand{\lapproxeq}{\lower
.7ex\hbox{$\;\stackrel{\textstyle <}{\sim}\;$}}

\newcommand{\qq}{$q\bar{q}$}
\newcommand{\cc}{$c\bar{c}$}

\newcommand{\ket}[1]{|           {#1}           \rangle}
\newcommand{\bra}[1]{\langle           {#1}           |}       
\def\3bar{$\bar {\hbox{\bf 3}}$}
  
\newcommand{\bc}{\begin{center}}  
\newcommand{\ec}{\end{center}}  
\newcommand{\btab}{\begin{tabular}}  
\newcommand{\etab}{\end{tabular}}

\newcommand{\ocleb}[6]{\langle{#1}  {#2}|{#3}  {#4};{#5}  {#6}\rangle} 
\newcommand{\tpn}{^3\mathrm{P}_0}
\newcommand{\tso}{^3\mathrm{S}_1}  
\newcommand{\grad}{\mathbf{\nabla}}
\newcommand{\sixj}[6]{ \left\{\begin{array}{ccc}  {#1} & {#2} & {#3} \\    {#4} 
& {#5} & {#6} \end{array}   \right\} }
\newcommand{\ninej}[9]{ \left\{\begin{array}{ccc}  {#1} & {#2} & {#3} \\    {#4} 
& {#5} & {#6}\\    {#7} & {#8} & {#9}  \end{array}   \right\} }
\newcommand{\rme}[3]{\bra{#1}|{#2}|\ket{#3}}

\newcommand{\ot}{\otimes}
\newcommand{\st}[1]{\Pi_{#1}}

\newcommand{\hatr}{\mathbf{\hat r}}

\newcommand{\uS}{\textrm{S}}
\newcommand{\uP}{\textrm{P}}
\newcommand{\uD}{\textrm{D}}
\newcommand{\uF}{\textrm{F}}
\newcommand{\spo}{^1\textrm{P}_1}
\newcommand{\tpo}{^3\textrm{P}_1}
\newcommand{\tpt}{^3\textrm{P}_2}

\newcommand{\tdo}{^3\textrm{D}_1}

\newcommand{\epem}{e^+e^-}

\def\q{\overline{q}}

\begin{document}

\title{\bf Dynamics of hadron strong production and decay}

\author{
T. J. Burns\footnote{e-mail: burns@thphys.ox.ac.uk}, F. E. Close\footnote{e-mail: F.Close1@physics.ox.ac.uk}
and C. E. Thomas\footnote{e-mail: C.Thomas1@physics.ox.ac.uk}}
%
\affiliation{ Rudolf Peierls Centre for Theoretical Physics,
University of Oxford, \\
1 Keble Rd., Oxford, OX1 3NP, United Kingdom}

\date{October 3, 2007}
\begin{abstract}
We generalize results of lattice QCD to determine the spin-dependent symmetries and factorization 
properties of meson 
production in OZI allowed processes. This explains some conjectures
previously made in the literature about axial meson decays and gives predictions for
exclusive decays of vector charmonia, including ways of establishing the structure of $Y(4260)$ 
and $Y(4325)$ from their S-wave decays. 
Factorization gives a
selection rule which forbids $e^+e^- \to D^* D_2$ near threshold with the tensor meson in 
helicity 2. The relations among amplitudes for double charmonia production
$\epem\to \psi\chi_{0,1,2}$ are expected to differ from the analagous
relations among light flavour production such as $\epem\to \omega f_{0,1,2}$.

\end{abstract}

\maketitle


\vskip 1.cm

\section{Introduction}

The dynamics of strong decay amplitudes are poorly understood. Definitive 
answers are not known to questions
as basic as: (i) are the \qq~ created in an OZI allowed decay spin singlet or 
spin triplet; (ii) what
is their overall $J^{PC}$; (iii) are the \qq~ created from the energy of the 
strong confinement field,
or from a hard gluon? It is our purpose in this paper to address these 
questions. We shall show that
results from lattice QCD imply that light \qq~ pair has spin 1 with 
an effective factorization
of constituent spin and orbital degrees of freedom such that the \qq~ pair in the 
initial meson are passive spectators.
By contrast, if heavy flavours are created, as in $e^+e^- \to \psi + \chi_J$, 
factorization is broken with
spin and momentum transferred from the initial \cc~ to the created pair, such as 
by a hard gluon. This
implies a radically different spin dependence of amplitudes, and of angular 
distributions, in analogue processes such as
$e^+e^- \to
\omega f_2$ relative to $e^+e^- \to \psi \chi_2$.

While Lattice QCD is now a mature guide for the masses of glueballs and hybrids, 
at least in the quenched approximation \cite{peardon,michael}, it is not yet mature enough to 
determine hadronic decays 
extensively. 
Consequently, at a fundamental level the
dynamics of such decays are not yet established.
Flux tube models of both spectra \cite{ip,bcs} and decays \cite{ikp,cp95} have 
been developed, in part stimulated by
attempts to model the lattice, and lattice work has confirmed their spectroscopy 
\cite{michael,bcs}.
The lattice is now beginning to confirm aspects of the flux tube model for some 
decays: 
specifically the lattice QCD studies of the
decays of hybrid $1^{-+} \to \pi b_1$ and $\pi f_1$ \cite{mm06} 
show quantitative features that were anticipated in flux-tube 
models \cite{ikp,cp95}, and in ref \cite{bc1} we 
showed that these approaches exhibit
remarkable agreement when compared under the same kinematic conditions. 
Specifically, for S-wave decay amplitudes at zero-recoil
the results are consistent with

\begin{equation}
a(\pi_1 \rightarrow \pi + b_1[^1\uP_1]) = 2 a(\pi_1 \rightarrow 
\pi + f_1[^3\uP_1]).
\end{equation}

\noindent where $\pi_1$ denotes the first gluonic excited hybrid with 
$J^{PC}=1^{-+}$.

In section II we describe the underlying assumptions of the
factorization hypothesis: (i) the hadrons' 
spins, $j$, separate
into two parts - the intrinsic spins of the constituents $ s$
and a residual component that transforms as angular momentum $l$, (ii)
the $l$ and $s$ degrees of freedom act independently throughout the 
transition (``factorization"), (iii)
the \qq~ pair produced has spin 1. 
In section IIA we demonstrate that the above ratio is immediate
within the factorization hypothesis. We identify further implications of factorization for the 
decays of axial and 
vector mesons in sections IIB and IIC;
the former confirms and explains a conjecture of \cite{abs} and the latter has 
implications for charmed meson production from a $\psi(\tso)$ initial state.
A helicity selection rule
is derived implying that in $\epem \to D^*D_2$ near threshold or from
a $\psi(^3\uS_1)$ initial state the $D_2$ cannot be produced with helicity two.
In section IID we apply the results of factorization to the decays of \cc~ to charmed mesons
near threshold in relative $S$-waves, and identify ways to distinguish
between hybrid and conventional interpretations of enigmatic $\psi$-like
states such as
$Y(4260)$ or $Y(4325)$. 

In section III we discuss application to $\epem$ annihilation for light and heavy flavours.
We show that decays triggered by hard gluons violate 
factorization, and that emerging data on $\epem \to \psi + \chi_j$
appear to support this. We propose tests for factorization and hard gluon production
mechanisms in $\epem \to \psi + \chi_j$ near threshold.


\noindent

\section { Factorization: Formulation and application}

Our approach to strong decays is similar in spirit to what was done in past decades
for electromagnetic and current induced transitions among hadrons, known 
variously as Melosh transformation 
or more generally ``single
quark transition algebra"\cite{melosh}. The empirically successful
hypothesis there was that the interaction of a current with 
a single quark triggers a transition, all other constituents being passive
spectators. That led to algebraic relations among amplitudes, which arose from 
the Clebsch-Gordan coefficients coupling the orbital, spin and total angular momentum 
projections
${l_z,s}_z$ to ${j}_z$ for the initial and final hadrons, and the ${l_z,s}_z$
algebraic transformation properties of the transition operators. 
While the relative strengths of the reduced matrix elements associated with each 
transition operator 
are in this general approach undetermined, the experimentally accessible range 
of helicity amplitudes for 
photo-excitation of
proton and neutron targets to different resonances within a supermultiplet led 
to experimentally testable 
relations among various amplitudes\cite{emamps}. Within the hypothesis of  $l,s$ 
factorization, analogous 
relations arise for strong decays. Specific models have implicitly assumed such 
factorization\cite{ley,ikp,cp95,bcps,ss}; we shall see
that the results from lattice QCD suggest that this property is realised in 
decays at least for light
flavours.

We consider the decay process of mesons $M \to M_1 + M_2$.  
In meson $M$ with spin $j$, the \qq~ have spin $s$ and residual angular momentum
$l$. We illustrate the structure of the amplitude for the particular case of a 
flux-tube that breaks to
 form a new \qq~ in a $^3\uP_0$ configuration leading to a pair of mesons $M_1$, 
$M_2$ with spins $j_1,j_2$
respectively, and their \qq~ having $s_1,l_1$, $s_2,l_2$. The generalisation 
will be immediate.

 The width for the decay of a meson into a pair of mesons involves a sum 
over 
 couplings of $j_1\otimes j_2$ to $j_{12}$ and relative partial waves $L$:
\be
\Gamma(slj\to s_1l_1j_1+s_2l_2j_2) \sim \sum_{j_{12}L}
\rme{(((s_1\ot  l_1)_{j_1}\ot(s_2\ot  l_2)_{j_2})_{j_{12}}\ot 
L)_j}{\mathbf{\sigma}\cdot\grad}{(s\ot l)_j}^2
\label{width}
\ee
\noindent where $\sigma$ transforms as a vector in spin-space and $\grad$ acts 
on the spatial
(orbital and radial) degrees of freedom.
Usually at this point specific wavefunctions are assumed and nonrelativistic 
expressions
calculated for the amplitudes \cite{ley,ikp,cp95,bcps,ss}. However, this introduces model 
dependence and 
obscures the 
more general underlying properties. Instead we shall factor the amplitude in such a way that the 
spin and space parts are 
separated \cite{tb1}, expressing all decay amplitudes as linear combinations of model-dependent 
spatial amplitudes, which in the present work are left general. 

The first step is to separate the spin and space degrees of freedom of the final 
state. The final state bra
\be
\bra{(((s_1\ot  l_1)_{j_1}\ot(s_2\ot  l_2)_{j_2})_{j_{12}}\ot L)_j}
\ee
is recoupled to form
states of good $(s_{12},l_{12},l_f)$:
\be
\bra{((s_1\ot s_2)_{s_{12}}\ot((  l_1\ot  l_2)_{l_{12}}\ot L)_{l_f})_{j}}
\ee
which involves a product of 6-$j$ and 9-$j$ coefficients. 
The spin and space parts then factorise and can be isolated. The result is

\begin{multline}
\rme{(((s_1\ot  l_1)_{j_1}\ot(s_2\ot  l_2)_{j_2})_{j_{12}}\ot 
L)_j}{\mathbf{\sigma}\cdot\grad}{(s\ot l)_j}=
\sum_{ s_{12}l_{12}  l_f} 	(-
)^{s+L+s_{12}+l_{12}+l_f}\st{l_fs_{12}l_{12}j_1j_2j_{12}}
\\
\ninej{s_1}{l_1}{j_1}{s_2}{l_2}{j_2}{s_{12}}{l_{12}}{j_{12}}
\sixj{s_{12}}{l_{12}}{j_{12}}{L}{j}{l_f}
\sixj{s_{12}}{s}{1}{l}{l_f}{j}
\rme{(s_1\ot s_2)_{s_{12}}}{\mathbf{\sigma}}{s}
\rme{(( l_1\ot  l_2)_{l_{12}}\ot L)_{l_f}}{\grad}{l}
\label{separatingspinandspace}
\end{multline}
with

\be
\st{ab\ldots}=\sqrt{(2a+1)(2b+1)\ldots}
\ee

\noindent The spin part is 
a  9-$j$ coefficient along with appropriate counting factors
\be
\rme{(s_1\ot s_2)_{s_{12}}}{\mathbf{\sigma}}{s}=
(-)^{s+s_1}
\st{1ss_1s_2s_{12}}\ninej{\frac{1}{2}}{\frac{1}{2}}{s_1}{\frac{1}{2}}{\frac{1}{2
}}{s_2}{s}{1}{s_{12}}.
\label{spinpart}
\ee
The 9-$j$ coefficient in eqn. \rf{spinpart} is zero for $s_1=s_2=s=0$; this is the well known 
spin singlet selection rule and is a consequence of the orthogonality of the spin wavefunctions. Note in the above a phase of $(-1)$ has been included for the permutation of quark and antiquark operators \cite{abs}, and the expression \rf{separatingspinandspace} is equivalent to that in Ref. \cite{bonnaz}.

The assumption driving the expansion of eqn. \rf{separatingspinandspace} is that the angular 
momentum and spin quantum numbers factorize and that the decay operator is overall scalar with a 
spin triplet part. The angular momentum algebra makes no reference to the spatial part of the 
operator, hence the linear combinations for $^3\uP_0$ and $^3\uS_1$ decay models, driven by operators 
$\sigma\cdot \grad$ and $\sigma\cdot \hatr$ respectively,  are the same. In a  $^3\uS_1$ model  the 
spatial contraction involves
$\hatr$ - the unit vector in the relative coordinate of the initial meson's \qq~. In constituent gluon models \cite{const} the spin-dependence of
the \qq~ creation again is via a $\sigma$ operator, while the spatial
contraction depends on the specific model wavefunctions. In general, for
any specific model there will be differing spatial dependence
but the overall spin coupling coefficients are identical.

Eqn. \rf{separatingspinandspace} expresses full decay amplitudes $a_{j_{12}L}$ as linear combinations of model 
dependent spatial amplitudes $A_{l_{12}Ll_f}$ of the form
\be
A_{l_{12}Ll_f}=\rme{(( l_1\ot  l_2)_{l_{12}}\ot L)_{l_f}}{\grad}{l}\label{spatialform}
\ee
The expansion applies to all  partial waves $L$  allowed by the conservation of angular momentum, including those which are parity-forbidden for a given set of spatial quantum numbers. It is the spatial matrix element itself which ensures the conservation of parity; this is verified in the expressions of Ref. \cite{tb1} where the spatial matrix elements for the production and decay of conventional and hybrid mesons are presented. Thus, for instance, decays of the type $^3\uP_1\to^1\uP_1~^1\uS_0$ are allowed in S-, P- and D-wave in general; if the initial $^3\uP_1$ is a conventional $1^{++}$ or hybrid $1^{+-}$ the S- and D-wave amplitudes vanish, whereas if the initial $^3\uP_1$ is a hybrid $1^{-+}$ the P-wave amplitude vanishes.

The approach taken here, however, is to exploit the relationships between the decay amplitudes $a_{j_{12}L}$ leaving the spatial amplitudes $A_{l_{12}Ll_f}$ undetermined.  These spatial amplitudes depend on the decay momentum and the spatial wavefunctions; thus, in the limit that the spatial wavefunctions of the mesons under comparison are the same and the momenta are the same, the expansion of Eqn. \rf{separatingspinandspace} relates decay amplitudes among families of states sharing the same spatial quantum numbers but having different spin and total angular momentum.  If for a given partial wave $L$ there is only one spatial matrix element of the form 
\rf{spatialform}, which we denote $A_L$, there are direct relations among amplitudes for states with different angular momentum quantum numbers. Three such cases are immediate:
\begin{enumerate}
\item one of the final states has orbital angular momentum zero ($l_2=0$) and the 
decay is in relative $S$-wave ($L=0$); thus $l_f=l_{12}=l_1$ and the amplitude is 
expressed in terms of a single matrix element $A_S$;
\item both final states have orbital angular momentum zero ($l_1=l_2=0$); thus 
$l_{12}=0$ and $l_f=L$ and the amplitude in a  partial wave $L$ can be expressed 
in terms of a single matrix element $A_L$;
\item the initial state and one of the final states have orbital angular momentum 
zero ($l=l_2=0$), thus $l_{12}=l_1$ and $l_f=1$ and the amplitude in a  partial wave 
$L$ can be expressed in terms of a single matrix element $A_L$.
\end{enumerate}

We now examine each of these three cases in turn with specific examples. 
In section \ref{eg1} the $S$-wave hybrid decays  $\pi_1\to b_1\pi$ and $\pi_1\to f_1\pi$ are 
shown to match results from lattice QCD and thereby to reveal significant information about the 
underlying dynamics (case 1). 
In section \ref{eg2}, $a_1\to\rho\pi$ and $b_1\to\omega\pi$ are examples of case 2; the analysis 
verifies a conjecture
that was made elsewhere \cite{abs} and establishes its origin. 
In section \ref{eg3} case 3 is applied to derive relations among decays of the type $^3\uS_1\to 
(^1\uP_1;^3\uP_j)+ (^1\uS_0;\tso) $.  A new selection rule is derived and the possibility of testing it 
in the context of $\epem$ annihilation producing flavoured and flavourless states is discussed. 
In section IID we
discuss ways of using these results to distinguish hybrid \cc~ from $\tso$ or $\tdo$ $\psi$ 
states.


\subsection {\bf $S$-wave decays of hybrid meson $\pi_1$}\label{eg1}

An immediate example of this factorization is the S-wave decays of the hybrid meson $\pi_1\to 
b_1\pi$ or $f_1\pi$.
The $\pi_1$ has $1^{-+}$ quantum numbers and $j=s=l=1$\cite{ikp,cp95}. In flux tube models and non relativistic constituent gluon
models\cite{const} the $l=1$ is explicit; in cavity and bag models it is
implicit in the definition of the TE gluon mode which transforms as $\hatr
\times \epsilon$ (e.g. see eq. (2.22) in \cite{bcm} and applied to hybrid
decays in \cite{bcd}). The final states have 
$l_1=j_1=1$ and 
$s_2=l_2=j_2=0$, differing in the spin quantum number $s_1=0$ ($b_1$) and $s_1=1$ ($f_1$). For 
decays in 
S-wave there is only one matrix element of the form (\ref{spatialform}), having $l_{12}=l_f=1$. 
In the 
summation over $l_{12}$ and $l_f$, this constraint is enforced by zeroes in the 9- and 6-$j$ 
coefficients
 which reduce to delta functions. This reduces the expansion of eqn. \rf{separatingspinandspace}
 to a simpler form and the amplitude for the initial state with spin $s_1$ is given by
\be
a_S(^3\uP_1\to^{s_1}\uP_1+^1\uS_0)=
\frac{3}{\sqrt 
2}\Pi_{s_1}\sixj{s_1}{1}{1}{1}{1}{1}\sixj{1/2}{1/2}{s_1}{1}{1}{1/2}\times A_S
\ee
where here $A_S$ is the spatial matrix element. Thus for the $^1\uP_1$ and $^3\uP_1$ modes
\bea
a_S(^3\uP_1\to^1\uP_1~^1\uS_0)&=&-\frac{A_S}{2\sqrt 3}\\
a_S(^3\uP_1\to^3\uP_1~^1\uS_0)&=&-\frac{A_S}{2\sqrt 6}
\eea
The flavour overlaps for $\pi_1\to b_1\pi$ and  $f_1(n\overline n)\pi$ are $\sqrt{2/ 3}$ and 
$\sqrt {1/3}$, so 
that the ratio of amplitudes is
\bea
\frac{a_S(\pi_1\to b_1\pi)}{a_S(\pi_1\to f_1\pi)}=2
\eea

\noindent which underwrites the result eq.(1) as found also in lattice QCD. The 
essential feature here is the
factorization of spin and space, and the assumption that the created \qq~ are 
spin-triplet. Note that the \qq~ creation with quark-spin 1 now appears explicitly in 9-$j$ and 
6-$j$ symbols; if the created pair has spin 0
the final 9-$j$ symbol for the $\pi b_1$ mode, equation \rf{spinpart}, has a zero in the bottom 
left corner, and since
\be
\ninej{\frac{1}{2}}{\frac{1}{2}}{0}{\frac{1}{2}}{\frac{1}{2}}{0}{0}{1}{0}  =0 
\ee\noindent the decay $\pi_1 \to \pi + b_1(^1\uP_1)$ would vanish. If	 spin is conserved, an 
initial state with $S=1$ can only decay to a pair of $S=0$ states if $S=1$ \qq~ 
is present, hence the
need for pair-creation to be spin-triplet for a non-zero amplitude.

Thus the results of lattice QCD, at least when applied to the decays of a hybrid 
meson\cite{mm06},
follow if the amplitude factors in space and spin, with the \qq~
pair creation being spin-triplet and an overall scalar. This does not 
distinguish $^3\uP_0$ from $^3\uS_1$ decay models.

\subsection{$S$ and $D$ wave decays of axial mesons}\label{eg2}

Ackleh et al\cite{abs} noted that the ratio of the $D/S$-wave amplitude ratios for $b_1 \to 
\omega \pi$ and $a_1 \to \rho \pi$
can be a sensitive discriminator among models. They found that if the \qq~ are 
created in the $^3\uP_0$ 
configuration, as commonly
assumed in flux-tube models, the ratio of $D/S$ ratios is

\be
\frac{\frac{a_D}{a_S}(a_1 \to \rho\pi)}{\frac{a_D}{a_S}(b_1 \to \omega\pi)} = -
\frac{1}{2}.
\label{theratio}\ee
\noindent They found the same ratio in the case of \qq~ creation by gluon exchange in the static 
limit (``colour coulomb")
but that it departs from $-1/2$ in the case of transverse gluon exchange. It was 
suggested that this might be useful as a
signature of the one-gluon exchange component in the physical decay amplitude, 
and noted that
experimentally the ratio is $-0.35 \pm 0.09$\cite{abs}, $2\sigma$ away from $-1/2$. Today the 
ratio 
\be
\frac{\frac{a_D}{a_S}(a_1 \to \rho\pi)}{\frac{a_D}{a_S}(b_1 \to \omega\pi)} = -
0.39 \pm 0.06
\label{exptratio}
\ee
has a greater precision\cite{pdg}  due to recent experiments\cite{experiments}, 
though the statistical deviation from $-1/2$ remains similar. Although the authors of ref 
\cite{abs} speculated
that the common ratio for $^3\uP_0$ and coulomb-gluon cases is because of a lack 
of spin-flip,
which is violated in the case of transverse gluon exchange and hence the 
deviation from $-1/2$ in that case,
they did not explicitly demonstrate the source.

In the factorisation scheme, the amplitude for these decays is proportional to a single matrix 
element; this is an example of case 2 cited above. Once again, zeroes in the Wigner coefficients 
reduce the expansion 
of eqn. \rf{separatingspinandspace} to a simpler form and enforce the conservation of angular 
momentum ($l_{12}=0$ 
and $l_f=L$), whereby the amplitude in a partial wave $L$ is proportional to a unique spatial 
matrix element $A_L$. The two
 decay modes of interest differ in the spin quantum number of the initial state, $s=0$ ($b_1$) 
and $s=1$ ($a_1$). 
 The amplitudes are given by
\be
a_L(^s\uP_1\to ^3\uS_1~^1\uS_0)=\frac{3}{\sqrt 
2}(-)^{L+1}\Pi_{s}\sixj{1}{s}{1}{1}{L}{1}\sixj{1/2}{1/2}{1}{1}{s}{1/2}\times A_L
\ee
This gives
\bea
&a_S(^1\uP_1\to ^3\uS_1~^1\uS_0)=-\frac{1}{2\sqrt 3}A_S\qquad&
a_D(^1\uP_1\to ^3\uS_1~^1\uS_0)=-\frac{1}{2\sqrt 3}A_D\\
&a_S(^3\uP_1\to ^3\uS_1~^1\uS_0)=-\frac{1}{ \sqrt 6}A_S\qquad&
a_D(^3\uP_1\to ^3\uS_1~^1\uS_0)= \frac{1}{2\sqrt 6}A_D
\eea
Thus we have established that the ratio eqn. \rf{theratio} is an immediate result of the 
factorization and \qq~ 
creation with spin 1.
A deviation from this ratio is indicative of a breaking of factorization, such 
as by a transverse gluon
which transfers spin and momentum (``spin-orbit coupling") in general. We shall 
return to this mechanism
for breaking of factorization in section III.

\subsection{$S$ and $D$ wave decays of vector mesons}\label{eg3}
If both of the vector states are  $^3\uS_1$ then the decay amplitudes are
$V+ (0^+,1^+,2^+)$, and the amplitudes in $S,D$- wave are each proportional to a unique spatial 
matrix element $A_S,A_D$; this is an example of case 3 discussed earlier.
Decays of vector mesons 
provide a range of tests 
of factorization and decay
dynamics. Decays of the type
\bea
^3\uS_1&\to&  ^3\uP_{0,1,2}+^3\uS_1\label{dec1}\\
^3\uS_1&\to&  ^3\uP_{1,2}+^1\uS_0\\
^3\uS_1&\to&  ^1\uP_1+^3\uS_1\\
^3\uS_1&\to&  ^1\uP_1+^1\uS_0\label{dec4}
\eea
all belong to the special case 3 described earlier; their decay amplitudes in a partial wave $L$ 
are each proportional to a unique matrix element $A_L$. Substituting into eqn. 
\rf{separatingspinandspace}  $l=0,s=j=1$ for the initial state and $l_1=1$ and $l_2=0$ for the 
final states gives the amplitude $a_{j_{12}L}$ for the decay in a partial wave $L$ with final 
states coupled to $j_{12}$:
\be
a_{j_{12},L}(^3\uS_1\to ^{s_1}\uP_{j_1}+s_2)
=
\sum_{s_{12}}(-)^{L+j_1+1}\st{1s_1s_2s_{12}s_{12}j_1j_{12}}
\sixj{s_1}{s_{12}}{s_2}{j_{12}}{j_1}{1}
\sixj{s_{12}}{1}{j_{12}}{L}{1}{1}
\ninej{\frac{1}{2}}{\frac{1}{2}}{s_1}{\frac{1}{2}}{\frac{1}{2}}{s_2}{1}{1}{s_{12}}
\times A_L
\label{gendec}\ee
The results are shown in Table \ref{tablesd} below. The pattern of amplitudes is realized in 
specific model calculations that have 
implicitly assumed factorization, 
e.g.\cite{bcps}, which give explicit expressions for the spatial 
dependences $A_{S(n)}$
and $A_{D(n)}$ for radial excitations $n$. The amplitudes in Table \ref{tablesd} differ from those in Ref. \cite{bcps} by a phase associated with the ordering of the angular momentum coupings.
\begin{table}[h]
\begin{tabular}{lrl}
\hline
$^3\uS_1\to ^3\uP_0~^3\uS_1$	\qquad &\qquad	$a_{1S}=$&$-A_S/2		$\\
                            	\qquad &	$a_{1D}=$&$0               	$\\
\hline
$^3\uS_1\to ^3\uP_1~^3\uS_1$	\qquad &	$a_{1S}=$&$A_S/\sqrt 3     	$\\
                            	\qquad &	$a_{1D}=$&$A_D/4\sqrt 3    	$\\
                            	\qquad &	$a_{2D}=$&$A_D/4           	$\\
\hline
$^3\uS_1\to ^3\uP_2~^3\uS_1$	\qquad &	$a_{1S}=$&$0               	$\\
				\qquad &	$a_{1D}=$&$A_D/4\sqrt 5    	$\\
				\qquad &	$a_{2D}=$&$-A_D/4\sqrt 3   	$\\
				\qquad &	$a_{3D}=$&$-A_D\sqrt{7/15} 	$\\
				\qquad &	$a_{3G}=$&$0               	$\\
\hline
$^3\uS_1\to ^1\uP_1~^3\uS_1$	\qquad &	$a_{1S}=$&$ A_S/ \sqrt 6   	$\\
				\qquad &	$a_{1D}=$&$-A_D/2 \sqrt 6  	$\\
				\qquad &	$a_{2D}=$&$ -A_D/2\sqrt 2  	$\\
\hline
$^3\uS_1\to ^1\uP_1~^1\uS_0$	\qquad &	$a_{1S}=$&$ A_S/2\sqrt 3	$\\
				\qquad &	$a_{1D}=$&$ A_D/2\sqrt 3   	$\\
\hline
$^3\uS_1\to ^3\uP_1~^1\uS_0$	\qquad &	$a_{1S}=$&$ A_S/\sqrt 6	$\\
				\qquad &	$a_{1D}=$&$ -A_S/ 2\sqrt 6 	$\\
\hline
$^3\uS_1\to ^3\uP_2~^1\uS_0$	\qquad &	$a_{2D}=$&$ A_D/ 2\sqrt 2  	$\\
\hline
\end{tabular}
\caption{Decay amplitudes $a_{j_{12}L}$ for the decays \rf{dec1}-\rf{dec4}}
\label{tablesd}
\end{table}

%

\vskip 0.1in

\subsubsection{{\bf $\psi(n\tso) \to$ flavoured mesons}}
The results of Table \ref{tablesd}
can be applied immediately to $\psi(n^3\uS_1) \to D_{0,2}D^*$ and also to $D_1D^{(*)}$. 
In the latter case data may be used to determine the mixing angle between $\spo$ and $\tpo$.

The eigenstates for axial flavoured mesons are in general mixtures of the
$^3\uP_1$ and $^1\uP_1$ states. Ref.\cite{cs06} defines the mixing angles by
\bea
|D_{1L}\rangle &=& \cos\phi |^1\uP_1 \rangle + \sin\phi |^3\uP_1 \rangle \nonumber \\
 |D_{1H}\rangle &=& - \sin\phi |^1\uP_1 \rangle + \cos\phi |^3\uP_1 \rangle 
\label{axialmix}
\eea
\noindent and discusses ways of determining them experimentally. The amplitudes for axial meson production as a function of mixing angle
follow from Table \ref{tablesd} with careful treatment of phase conventions for the spin-mixed states.  Ref.\ \cite{newheavymesons} adopted the following conventions: for $q\bar c$ states (as opposed to $c\bar q$ states) and with orbital and spin angular momentum combined in the order $(l\ot s)_j$, the heavy quark limit gives $\phi = -54.7^o$ \cite{newheavymesons} so that the states are
\bea
|\bar D_{1L}\rangle = \sqrt{\frac{1}{3}}|^1\uP_1 \rangle - \sqrt{\frac{2}{3}}|^3\uP_1 \rangle \nonumber \\
|\bar D_{1H}\rangle = \sqrt{\frac{2}{3}}|^1\uP_1 \rangle +\sqrt{\frac{1}{3}} |^3\uP_1 \rangle 
\label{axialmixheavy}
\eea
\noindent
The amplitudes of eqn. \rf{separatingspinandspace}, shown in Table \ref{tablesd}, are for the topology in which the created  $q$ $(\bar q)$ ends up in the meson with quantum numbers $s_1l_1j_1$ $(s_2l_2j_2)$. If the axial states are labelled with the quantum numbers $s_1l_1j_1$ they are $q\bar c$ states in accordance with conventions of ref \cite{newheavymesons}.  However, the conventions in the present paper are that amplitudes apply to meson spin coupling in the order $(s\ot l)_j$, so there is relative minus sign associated with the $^3\uP_1$ part of the amplitude. Thus for the mixed $\bar D_{1H},\bar D_{1L}$ states in the heavy quark limit, the amplitudes for $^3\uS_1\to \bar D_{1L}D,\bar D_{1H}D$ are
\bea
a_{j_{12}L}(^3\uS_1\to \bar D_{1L}D)&=&
\sqrt{\frac{1}{3}}a_{j_{12}L}(^3\uS_1\to ^1\uP_1~^1\uS_0)
+\sqrt{\frac{2}{3}}a_{j_{12}L}(^3\uS_1\to ^3\uP_1~^1\uS_0),\label{D1LD}\\
a_{j_{12}L}(^3\uS_1\to \bar D_{1H}D)&=&
\sqrt{\frac{2}{3}}a_{j_{12}L}(^3\uS_1\to ^1\uP_1~^1\uS_0)
-\sqrt{\frac{1}{3}}a_{j_{12}L}(^3\uS_1\to ^3\uP_1~^1\uS_0),\label{D1HD}
\eea
and likewise for $^3\uS_1\to \bar D_{1L}D^*,\bar D_{1H}D^*$. This gives the relative decay widths 
(up to phase space corrections) shown in Table \ref{tablecharm} below.

\begin{table}[h]
\begin{tabular}{l rl rl }
\hline
				
			&$S^2$&	&$D^2 $&\\
\hline
$D_0D^*$		&1&	&0&		\\	
$D_{1L}D^*	$	&2&        	&0   &           	       		\\
$D_{1H}D^*	$	&0&        	&1   &           	       		\\
$D_2D^*$		&0   &         	&2    &            	        		\\
\hline
$D_{1L}D$   &1&    &0&   \\
$D_{1H}D$   &0&    & $\frac{1}{2}$&   \\
$D_2D$	   &0&    & $\frac{1}{2}$&   \\
\hline
\end{tabular}
\caption{Relative widths $\tso \to D^*D_{0,1,2}$ or $D D_{0,1,2}$; the states
$D_{1L,H}$ are light and heavy axial mesons in the heavy quark limit.}
\label{tablecharm}
\end{table}

Hence in the heavy quark limit
\bea
\Gamma(\psi(n\tso)\to D^*D_{1L}) = 2\Gamma(\psi(n\tso) \to DD_{1L}) \\
\Gamma(\psi(n\tso)\to D^*D_{1H}) = 2\Gamma(\psi(n\tso)\to DD_{1H})
\eea
\noindent as well as
\bea
\Gamma(\psi(n\tso) \to D^*D_{1L}) & = & 2\Gamma(\psi(n\tso) \to D^*D_{0})\\
\Gamma(\psi(n\tso) \to D^*D_{1H}) &= &\frac{1}{2}\Gamma(\psi(n\tso) \to D^*D_{2})\\
\Gamma(\psi(n\tso) \to DD_{1H}) &= &\Gamma(\psi(n\tso) \to DD_{2})
\eea
In addition there is a selection rule that the $D_2$ is produced only in helicity 0 or 1; {\it 
i.e} 
denoting helicity states by $0,(\pm),(\pm\pm)$ then

\be
a(\psi(n\tso) \to \bar D_2(\pm\pm) D^*(\mp))=0
\ee
\noindent This will be derived in the next section.


\subsubsection{\bf Helicity selection rule}

In the factorization scheme, the decay of a transversely polarised $\tso \to ^3\uP_2+^3\uS_1$, 
with the
tensor meson maximally polarised along the decay axis, is predicted to vanish:
\be
a(^3\uS_1(+)\to ^3\uP_2(++)^3\uS_1(-))=0\label{helrule}
\ee


This selection rule is a test of factorization; a significant non-zero strength 
for this helicity amplitude in a decay $1^{--} \to 1^{--}2^{++}$
signals either a breakdown of factorization or the presence of $^3\uD_1$ in $1^{--}$ or of $^3\uF_2$
in $2^{++}$.  
The origin of the selection rule is most transparent if we consider the 
helicity amplitude structure
directly. Its generality can then be assessed by transforming to partial wave amplitudes.

\begin{figure}[h]
\setlength{\unitlength}{0.00023300in}%
\begingroup\makeatletter\ifx\SetFigFont\undefined
\def\x#1#2#3#4#5#6#7\relax{\def\x{#1#2#3#4#5#6}}%
\expandafter\x\fmtname xxxxxx\relax \def\y{splain}%
\ifx\x\y   
\gdef\SetFigFont#1#2#3{%
  \ifnum #1<17\tiny\else \ifnum #1<20\small\else
  \ifnum #1<24\normalsize\else \ifnum #1<29\large\else
  \ifnum #1<34\Large\else \ifnum #1<41\LARGE\else
     \huge\fi\fi\fi\fi\fi\fi
  \csname #3\endcsname}%
\else
\gdef\SetFigFont#1#2#3{\begingroup
  \count@#1\relax \ifnum 25<\count@\count@25\fi
  \def\x{\endgroup\@setsize\SetFigFont{#2pt}}%
  \expandafter\x
    \csname \romannumeral\the\count@ pt\expandafter\endcsname
    \csname @\romannumeral\the\count@ pt\endcsname
  \csname #3\endcsname}%
\fi
\fi\endgroup
\begin{picture}(5788,4588)(1757,-12705)
\thicklines
\put(5401,-9661){\vector( 1, 1){1050}}
\put(6676,-12436){\vector(-1, 1){975}}
\put(5701,-10411){\vector( 1,-1){975}}
\put(7501,-8611){\vector(-1,-1){1050}}
\put(5101,-10861){\vector(-1, 0){2400}}
\put(1801,-9961){\vector( 1, 0){1500}}
\put(5701,-10411){\line( 1, 1){1800}}
\put(5701,-10411){\makebox(6.6667,10.0000){\SetFigFont{10}{12}{rm}.}}
\put(5701,-10411){\line( 1,-1){1800}}
\put(1801,-10861){\line( 1, 0){3300}}
\put(5101,-10861){\line( 1,-1){1800}}
\put(1801,-9961){\line( 1, 0){3300}}
\put(5101,-9961){\line( 1, 1){1800}}
\put(1801,-10861){\line( 1, 0){3300}}
\put(5101,-10861){\line( 1,-1){1800}}
\put(5101,-10861){\vector(-1, 0){2400}}
\put(1801,-9961){\line( 1, 0){3300}}
\put(5101,-9961){\line( 1, 1){1800}}
\put(5701,-10411){\vector( 1,-1){975}}
\put(7501,-8611){\vector(-1,-1){1050}}
\put(1801,-9961){\vector( 1, 0){1500}}
\put(5701,-10411){\line( 1, 1){1800}}
\put(5701,-10411){\makebox(6.6667,10.0000){\SetFigFont{10}{12}{rm}.}}
\put(5701,-10411){\line( 1,-1){1800}}
\put(900,-11000){$\bar q_4$}
\put(900,-10000){$q_1$}
\put(7500,-8500){$\bar q_2$}
\put(7000,-8000){$q_1$}
\put(7700,-12500){$q_3$}
\put(7200,-13000){$\bar q_4$}
\end{picture}
\caption{Strong decay topology.}
\label{topology}
\end{figure}
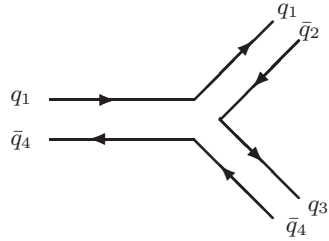

First consider the helicity picture. The decay is 
\be
q_1\bar{q}_4 \to [q_1\bar{q}_2] + [q_3\bar{q}_4]
\ee 
through the 
creation of $\bar{q}_2q_3$ (Fig \ref{topology}).
Denoting fermions with $S_z= \pm 1/2$ by $u,d$ respectively, the initial $^3\uS_1(+)$ has 
its \qq~ spins oriented $u_1u_4$. 
The final $^3\uP_2(++)^3\uS_1(-)$ then has to be $[u_1\bar{u}_2]+[d_3\bar{d}_4]$ with the $[u_1\bar{u}_2]$ also having $L_z = +1$, so spin-flip is required for a non-vanishing amplitude.

Spin conservation on spectator lines following the steps above,
or the diagrammatic techniques of ref. \cite{abs}, enable
relations among helicity amplitudes to be calculated in such factorizing models. 


In order to expose the more general dynamics underpinning this selection rule,
and to exhibit the relations among the various helicity amplitudes,
it is useful to transform
between helicity and partial wave amplitudes. 
As before, consider a state with spin $j$ decaying to two 
particles with spins respectively $j_1,j_2$. The final state can be characterised by quantum numbers $(j_{12},L)$ or by helicity quantum numbers $(\lambda_1,\lambda_2)$; the translation between the two bases, for an initial state with spin projection
$m$ along some axis, is given by
\begin{equation}
|jm;\lambda_1,\lambda_2 \rangle = \sum_{L j_{12}}\sqrt{\frac{2L+1}{2j+1}}  \langle 
j_{12}\lambda | j_1 \lambda_1; j_2 -\lambda_2 \rangle 
\langle j\lambda |j_{12}\lambda; L0 \rangle |jm; j_{12} L \rangle
\label{helpwave}
\end{equation}
This enables helicity amplitudes $a_{\lambda_1\lambda_2}$ to be written as linear combinations of partial 
wave amplitudes $a_{j_{12}L}$. We are interested here in decays of the type
\be
V(\lambda)\to \chi_{j_1}(\lambda_1)+V(-\lambda_2)\label{generic}
\ee
with 
$\lambda_1 - \lambda_2 =\lambda$, the relation between helicity and partial wave amplitudes follows from \rf{helpwave} above with $j=j_2=1$,
\be
a_{\lambda_1\lambda_2}=
\sqrt{\frac{1}{3}}\ocleb{1}{\lambda}{j_1}{\lambda_1}{1}{-\lambda_2}a_{1S}+
\sqrt{\frac{5}{3}}\sum_{j_{12}}\ocleb{1}{\lambda}{j_{12}}{\lambda}{2}{0}\ocleb{j_{12}}{\lambda}{j_1}{\lambda_1}{1}{-\lambda_2}a_{j_{12}D}+
\sqrt{ 3         }             \ocleb{1}{\lambda}{3     }{\lambda}{4}{0}\ocleb{3     }{\lambda}{j_1}{\lambda_1}{1}{-\lambda_2}a_{3     G}
\ee
The resulting relations are shown in the first column of Table \ref{helmult}. These relations apply generically to the decay of any vector meson to any scalar, axial or tensor meson $\chi_{j_1}$. For decays of the type
\be
^3\uS_1(\lambda)\to ^3\uP_{j_1}(\lambda_1)+^3\uS_1(-\lambda_2),\label{specific}
\ee
where each of the vectors are explicitly in a  $^3\uS_1$ state and the $\chi_{j_1}$ is a $^3\uP_{j_1}$ state, the amplitude is obtained by substituting for $a_{j_{12}L}(^3\uS_1\to ^3\uP_{j_1}~^3\uS_1)$ from Table \ref{tablesd}; the results are shown in the second column of Table \ref{helmult}. The selection rule \rf{helrule} is explicit in the last line of  Table \ref{helmult} and follows immediately substituting
\be
a_{1S}=a_{3G}=0;\qquad a_{1D}=A_D/4\sqrt 5;\qquad a_{2D}=-A_D/4\sqrt 3;\qquad a_{3D}=-A_D\sqrt {7/15};
\ee


\begin{table}[h]
\begin{tabular}{lr ll  }
\hline
				&		&$V(\lambda)\to \chi_{j_1}(\lambda_1)+V(-\lambda_2)$		&$^3\uS_1(\lambda)\to ^3\uP_{j_1}(\lambda_1)+^3\uS_1(-\lambda_2)$\\
\hline
$j_1=0			$	&$a_{0 , 0}$ 	&$=a_{1S}/\sqrt{3} -a_{1D}\sqrt{2/3}$		 		&$=-A_S/2\sqrt{3}$\\
				&$a_{0 , +}$ 	&$=a_{1S}/\sqrt{3} + a_{1D}/\sqrt{6}$		 		&$=-A_S/2\sqrt{3}$	\\	
\hline
$j_1=1$	&$a_{0 , 0}$ 	&$=0$						 		&$=0$\\
				&$a_{+ , 0}$ 	& $=a_{1S}/\sqrt{6} + a_{1D}/\sqrt{12} -a_{2D}/2$ 	        &$=A_S/3\sqrt{2} - A_D/12$\\
				&$a_{0 , +}$ 	&$=-a_{1S}/\sqrt{6} - a_{1D}/\sqrt{12} -a_{2D}/2$	   	&$=-A_S/3\sqrt{2} - A_D/6$    \\
				&$a_{+ , -}$ 	&$=a_{1S}/\sqrt{6} - a_{1D}/\sqrt{3} $ 		         	&$=A_S/3\sqrt{2} - A_D/12$    \\
\hline
$j_1=2$			&$a_{0 , 0}$ 	&$=-a_{1S}\sqrt{2/15} +2 a_{1D}/\sqrt{15} +3 a_{3D}/\sqrt{35} -2a_{3G}\sqrt{3/35}$   
									     	     	&$=-A_D/2\sqrt{3}$    \\
				&$a_{+ , 0}$ 	&$=-a_{1S}/\sqrt{10} - a_{1D}/\sqrt{20} -a_{2D}/\sqrt{12} +4a_{3D}/\sqrt{105} +2a_{3G}/\sqrt{35}$   
									          	&$=-A_D/4$    \\
				&$a_{0 , +}$ 	& $=a_{1S}/\sqrt{30} + a_{1D}/\sqrt{60} +a_{2D}/2 +2a_{3D}/\sqrt{35} + a_{3G}\sqrt{3/35}$ 		
									         	& $=-A_D/2\sqrt{3}$     \\
				&$a_{+ , -}$ 	&$=a_{1S}/\sqrt{10} - a_{1D}/\sqrt{5} + a_{3D}\sqrt{3/35} - 2a_{3G}/\sqrt{35}$   
									          	&$=-A_D/4$    \\
				&$a_{++,- }$ 	&$=a_{1S}/\sqrt{5} + a_{1D}/\sqrt{10} -a_{2D}/\sqrt{6} +a_{3D}\sqrt{2/105} +a_{3G}/\sqrt{70}$   
									          	&$=0 $    \\
\hline
\end{tabular}
\caption{Column 1 expresses helicity amplitudes $a_{\lambda_1\lambda_2}$ in terms of partial wave amplitudes   $a_{j_{12}L}$ for decays \rf{generic} of generic vector states. Column 2 expresses helicity amplitudes in terms of spatial amplitudes $A_L$ for decays \rf{specific} with explicit $s$ and $l$ quantum numbers.}
\label{helmult}
\end{table}





The amplitude $a_{3G}\equiv 0$ in $^3\uP_0$ or $^3\uS_1$ models since $(l_{12} = l_2 = 1) \otimes (L=3)$ can couple to $(l_f = 2,3,4)$,
which cannot couple to the $l=0$ initial state by the vector decay operators $\grad$ or $\hatr$. The appearance of this zero can be tested by measurement of the various helicity amplitudes which satisfy the linear relation 
\begin{equation}
a_{0,0} -2/\sqrt{3}a_{+,0} + 2/\sqrt{3} a_{+,-} = a_{0,+} + 1/\sqrt{6} a_{++,-}
\label{linear}
\end{equation}


 The selection rule $V(+) \to T(++)V(-) = 0$ can be violated
by failure of factorization, such as when single gluon exchange produces the 
$\bar{q}_2q_3$ and flips-spin
such as $u_1 \to d_1$, or if there are $^3\uD_1$ admixtures in the
wavefunctions of the produced or initiating vector mesons. 
The general property that breaks factorization and mixes $^3\uD_1$ components in the produced
vector meson is 
essentially the
same: in models the latter
is generated by spin-orbit coupling, such as from gluon exchange\cite{mixing}.
To the extent that vector mesons and $e^+e^-$ annihilation are dominated by 
$^3\uS_1$ configurations,
and the strong decay amplitude factorizes,
the selection rule will apply. For the $\psi(4415)$, which is consistent with being 
$4^3\uS_1$\cite{bgs},
the decays $\psi(4415) \to DD_2$ 
have been observed by initial state radiation \cite{4415}; 
our selection rule may be testable on the high mass side of the $\psi(4415)$
in its decays to the low mass tails of $D^*(2010)$
and $D_2(2460)$ respectively. It can also be tested in the $\epem$ continuum immediately 
around 4.5GeV as $\tdo$ contamination is expected to be minimal\cite{bgs}.


\subsection{{\bf Hybrid and Exotic Charmonium}}\label{eg4}
\vskip 0.1in


While $\tso$ $\psi$ states are expected to dominate the couplings to $\epem$ annihilation,
there are local \cc~ resonance structures in in the 4-5GeV energy range whose structure is
still unestablished\cite{bgs}. In particular there are the enigmatic structures $Y(4260)$ and
$Y(4325)$\cite{4260}. These  have no natural assignment within \cc~ spectroscopy and explanations
include hybrid charmonium, or molecules (e.g. by either $cq\bar{c}\bar{q}$
tetraquarks or
$DD_1$ and $D^*D_0$ attractive forces via $\pi$ exchange), or even effects associated with S-wave 
charmed meson thresholds\cite{fecbled}.

These states are near to the S-wave thresholds for $DD_1, D^*D_{0,1,2}$.  Such decays are an 
example of case 1 discussed earlier: each S-wave amplitude is proportional to a single spatial 
matrix element. The coefficient is a function of the spin and orbital angular momentum of the 
vector initial state, and thus the pattern among decay amplitudes differs for $^3\uS_1,^3\uD_1$ and hybrid interpretations, where for the latter the \cc~ have $l=1$ and $s=0$. Eqn. \rf{separatingspinandspace} gives the coefficient of the spatial matrix element, and the results are shown  in Table \ref{charmonium}. For axial mesons the amplitudes are shown in both the 
$^1\uP_1$-$^3\uP_1$ basis and in the heavy quark limit, where for the latter the amplitudes are given by eqns. \rf{D1LD} and \rf{D1HD} and their analogues.


\begin{table}[h]
\begin{tabular}{l rl rl rl}
\hline
				
			&$\tso$		&	&$\tdo $		&  	& $ ^1\Pi\uP_1$&\\
\hline
$D_0D^*$		&$-1/2$		&	&0			&	& $1/3\sqrt{2}$&\\	
$D_1(^1\uP_1)D^*$	&$1/\sqrt{6}$ 	&       &$-1/2\sqrt{6} $   	& 	& $1/2\sqrt{3}$& \\
$D_1(^3\uP_1)D^*$	&$1/\sqrt{3}$ 	&       &$1/4\sqrt{3} $   	& 	& 1/2$\sqrt{6}$ &\\
$D_2D^*	$		&0   		&       &$1/4\sqrt{5}   $ 	& 	& $-\frac{1}{6}\sqrt{\frac{5}{2}} $&\\
$D_1(^1\uP_1)D$		&$1/2\sqrt{3}$	&	&$1/2\sqrt{3}$		&	& 0 &	\\	
$D_1(^3\uP_1)D$		&$1/\sqrt{6} $	&       &$-1/2\sqrt{6}$    	&  	& $-1/2\sqrt{3} $&\\
$D_{1L}D	$	&$ 1/2 $	&       &$0$    		&  	& $-1/3\sqrt{2} $&\\
$D_{1H}D	$	&$ 0 $		&       &$1/2\sqrt{2}$   	 &  	&$ 1/6$&\\
\hline
\end{tabular}
\caption{Relative $S$  wave amplitudes for vector charmonia decays with $\tso$,$\tdo $	 and $ ^1\Pi\uP_1$ (hybrid) configurations; the states $D_{1L},D_{1H}$
refer to axial mesons in the heavy quark limit. 
}
\label{charmonium}
\end{table}
If production of charmed mesons in the decays of 
$\tso$ \cc~ is confirmed to factorise, then using Table \ref{charmonium} the relative decay 
amplitudes to $DD_1, D^*D_{0,1,2}$  may be used to determine the
structure of \cc~ states that are near to the S-wave thresholds. In particular this applies to 
$Y(4260)$ and $Y(4325)$. There are characteristic zeroes that may occur
for vector meson decays:
\bea
&\Gamma(^3\uS_1 \to D_{1H}D) &= 0\label{firstreln}\\
&\Gamma( ^3\uD_1 \to D_{1L}D) &= 0\\
&\Gamma(^1\Pi\uP_1\textrm{(hybrid)}] \to D_1(^1\uP_1)D) &= 0
\label{zeroes}
\eea
 The first pair of zeroes arise from the affinity of light and heavy $D_{1L},D_{1H}$ for $S$ and 
$D$ couplings
respectively, and the zero \rf{firstreln} was noted by ref. \cite{bgs}. For the hybrid decay the result follows from the conclusion of lattice
QCD, section IIA, that decays are driven by \qq~ creation in spin-triplet, which 
implies that a pair of spin-singlets (such as $D$ and $^1\uP_1$) cannot be produced from
a spin-singlet, such as a hybrid vector \cc~ . In practice these predictions will be
affected by mixing, which can be determined from other processes
(e.g. see \cite{cs06}), and by phase space. The relative rates are
insensitive to form factor effects at low momenta (see for example refs \cite{ley,bcps,cs06}). 

\vskip 0.1in
\section{{\bf Electron-positron annihilation}}

 We consider now the production of meson pairs in $\epem$ annihilation,
 supposing that such processes proceed through the strong decay of a
virtual
 quarkonia state \be \epem\to q_1\q_4 \to q_1\q_2 + q_3\q_4. \ee An
analagous
 model for $\psi$ decays to light flavour meson pairs was found to be
consistent
 with data assuming the $q_1\q_4$ state is some radial excitation
$n~^3\uS_1$
 \cite{missingref}. If the same applies here, the relative production
amplitudes
 of $^1\uP_1~^1\uS_0$, $^1\uP_1~^3\uS_1$, $^3\uP_j~^1\uS_0$ and
 $^3\uP_j~^3\uS_1$ will have the pattern of Table I, independently of $n$.
Such
 relations apply in the limit of equal momentum decays and provided there
is not
 an unfortunate double conspiracy in which both a single $n$ dominates and
a
 node in its amplitude coincides with the kinematic region of interest.
This
may be checked by varying $q^2$ to see if the ratios are stable or vary in an oscillatory
or nodal manner. The assumption that the pair $q_1\q_4$ is dominantly in a $^3\uS_1$
configuration is reasonable above charm threshold where
the coupling $\epem \to \tdo$(\cc~) is theoretically and empirically suppressed\cite{pdg,bgs}. 
The results of Table \ref{tablesd} then apply immediately to $\epem \to D^*D_{0,2}$ and 
also to $D^{(*)}D_1$, which in the latter case may be used to determine the mixing angle between 
$\spo$ and $\tpo$. 
While this is strictly true on a $^3\uS_1$ $\psi$ resonance, it may also be expected to
hold through the 4-5 GeV region of interest where $^3\uS_1$ is predicted to dominate the $\epem$ 
cross section.

Application to $\epem \to \psi + \chi_j$ also follows if this is dominated by strong flux-tube
formation and breaking. We shall argue in section IIIB that this is more likely to be dominated 
by
(perturbative) gluon exchange, which breaks factorization and 
gives a different pattern of amplitudes than 
strong flux-tube breaking. Our results may be used to test this hypothesis.

In the case of light flavours the neglect of $\epem \to \tdo$(\qq~) is more questionable. The
leptonic widths of $\tdo$(\qq~) are nonetheless expected to be relatively small\cite{gi85}, and 
empirically
the known vector mesons appear to fit well with (radially excited) $\tso$ 
with some mixing with hybrid vectors without need for significant $\tdo$\cite{cdsbook}. 
This is clearly an area whose phenomenology merits further clarification. To that end we 
apply our results with $\tso$ dominance to light flavours in the hope of shedding further 
light on this sector and isolating $\tdo$ states. 
For $\psi$ decays this simple assumption appears to be consistent with existing 
data\cite{pdg,bc2}.

\subsection{{\bf $\epem \to $ flavourless mesons} }

In the case of $\epem \to $neutral states, charge conjugation restricts the production of 
axial-vector mesons
in association with $0^{-+}$ or $1^{--}$
to 
$\epem \to V+ ^3\uP_1$ or $^1\uS_0 + ^1\uP_1$.
The amplitudes of Table \ref{tablesd} apply and 
the relative rates then follow by application of equation \rf{separatingspinandspace}:
 

\be
^3\uS_1+\tpn~:~^3\uS_1+\tpo~:~^3\uS_1+\tpt~:~^1\uS_0 + ^1\uP_1 ~=~ 3S^2~:~4S^2+D^2~:~6D^2~:~S^2+D^2
\ee
\noindent and hence

\be
\sigma(^3\uS_1+\tpo) =\frac{4}{3}\sigma(^3\uS_1+\tpn) + \frac{1}{6}\sigma(^3\uS_1+\tpt)
\label{sumrule1}
\ee
\noindent together with
\be
3\sigma(^1\uS_0 + ^1\uP_1) =\sigma(^3\uS_1+\tpn) + \frac{1}{2}\sigma(^3\uS_1+\tpt)
\label{sumrule2}
\ee
\noindent and their corrolary
\be
\sigma(^1\uS_0 + ^1\uP_1) =\frac{1}{8}\sigma(^3\uS_1+^3\uP_2) + \frac{1}{4}\sigma(^3\uS_1+^3\uP_1).
\label{sumrule3}
\ee
\noindent Note that necessarily

\be
\sigma(\epem \to ^3\uS_1+\tpo) > \sigma(\epem \to ^3\uS_1+\tpn).
\label{smallscalar}
\ee
\noindent For flavoured states the two axial mesons are mixtures of
$^1\uP_1$ and $^3\uP_1$; whatever the mixing angle may be, the inequality holds true in the sense 
that
the $^3\uS_1+^3\uP_0$ production rate cannot exceed those of both of axial mesons. 
In the case of charge conjugation eigenstates we are restricted to applying it to
light flavours or to $\epem \to \psi + \chi_j$. The former case is less well controlled 
theoretically, due to relativistic effects and potential contamination from $\tdo$ background in 
$\epem$ annihilation,
though the above relations appear to be consistent with data and are discussed in ref.\cite{bc2}. 
One of the central applications of the present paper will be to test
these predictions against data on $\epem \to \psi + \chi_j$ where preliminary
indications are that the relation eq.\rf{smallscalar} 
is violated\cite{belleddstar}. This is discussed in section IIIB.


The amplitude $V(-)T(++)$, where here $T$ denotes a tensor meson, should also be measured for light flavours where $VT$ modes are 
prominent, especially in
$\psi$ decay. Within the factorization hypothesis and $\tso$ dominance
 the $V f_2$ cannot be produced with $f_2$ maximally polarised;
$a[e^+e^- \to V(j_z = -1)f_2(j_z=+2)] = 0$ .
This may be studied in $\epem  \to 5\pi = 2\pi^+2\pi^-\pi^0$ by isolating the channel 
$\omega f_2$;
the $\omega$ being a narrow state can enable the angular distribution in decay $f_2 
\to \pi^+\pi^-$ to be measured.
The main background here is the potential contamination from
$\epem$ annihilation in the $\tdo$ state. Although models and data do not suggest
this is significant, nonetheless one cannot rule it out. If the amplitude is empirically found
to be small, in accord with the selection rule, one could turn this to advantage and
study the amplitude as a function of $q^2$ and observe if it turns on in the neighbourhood of the predicted
$^3\uD_1$ resonances around 2.2GeV\cite{gi85}.


\subsection{{\bf Factorization breakdown and preformation by OgE}}

Data on $e^+e^- \to \psi + X$ at 10.6 GeV c.m. energy show
three prominent enhancements
$X$ in $e^+e^- \to \psi + X$~\cite{belleddstar}, which are
consistent with being the $\eta_c,\eta_c'$ and $\chi_0$. The observed pattern of states 
appears radically
different to what is seen for light flavours, for example 
the apparent prominence of $e^+e^- \to \psi + \chi_0$ with only a hint of 
$\chi_1$ and much suppressed $\chi_2$ contrasts with light flavours where $e^+e^- \to \omega f_2$ 
is
clearly seen\cite{pdg}.
This suggests that this process for heavy flavours
may be controlled by a production mechanism where factorization is broken.

On theoretical grounds one expects that strong factorization may be overwritten here.
In $e^+e^-$ annihilation at $E >$ 6 GeV, creation of an initial \cc~ leaves up 
to 3 GeV available.
As the \cc~ separate, forming a strong flux tube up to $O(1 fm)$ long, the 
energy of $O(1 $GeV)
enables light-flavoured \qq~ to form. That is the familiar
dynamics that appears to be realized at low energies for light flavours. In the 
present example,
the most probable circumstance is that the excess energy produces multiple \qq~ 
leading to final states 
$D\bar{D}\pi\pi...$. 
The experimental selection on final states $\psi X(c\bar{c})$ isolates an 
unlikely configuration
where the 3 GeV has produced a \cc~ exclusively. For the flux tube to grow 
without splitting until it contains 3 GeV of energy
would require it to extend to distances exceeding $\Lambda^{-1}_{QCD}$.  This is 
exponentially unlikely
with increasing energy.

Alternatively the energy can be transmitted through a single gluon which converts 
to \cc~. While this is
perturbative and expected to be sub-dominant for processes involving light 
flavour creation, the question arises 
at what energy or for what flavours this dominates over flux-tube breaking. The 
purpose of this section is to propose ways of
answering this by experiment. We make specific reference to $e^+e^- \to 
(c\bar{c}) + (c\bar{c})$
as there are emerging data in the form of $e^+e^- \to \psi + X$.

As momentum flows through the gluon, it can transfer spin or angular momentum 
between the \cc~ to which
it is coupled. In general therefore we anticipate that factorization will break 
down.

Ref \cite{abs} have considered these matrix elements in the explicit non-relativistic limit 
- (see Appendix B of ref \cite{abs}, especially eqs B5-B7). In that limit the gluon-exchange
operation transforms as ${\bf S.S}$ and ${\bf L.S}$   but there is no ${\bf S.L}$
operator
(where the first operator refers to the transformation
property of the gluon emission and the second operator to that of \qq~ creation). 
Thus in the strict 
non-relativistic limit of that model, the $V(-)T(++)$ selection rule would appear to survive
for the decay of a $\tso$ vector meson. 
This is no surprise following the discussion after eq.\rf{helrule} : non-zero amplitude requires 
spin flip at the 
emission vertex and orbital flip at the $c\bar{c}$ creation vertex; while the former occurs
in the non-relativistic limit, the latter does not. 

However, in $\epem$ annihilation at $q^2 \equiv E^2_{c.m}$, the production of a
\cc~ allows an  ${\bf S.L}$ 
operator at $O(q^2/m^2_c)$ . 
An explicit calculation of the gluon exchange contributions to $e^+e^- \to \psi + \chi_j$ has 
been made in NRQCD in ref \cite{braaten} and a non-
vanishing amplitude for
$V(-)T(++)$ is found even at threshold, in accord with the discussion above.
Threshold is when $q^2 = 16m^2_c$;
the amplitudes depend upon $r^2 \equiv 16m^2_c/q^2$. At high energies, where $r^2 \to 0$
the contribution from $\epem \to \tdo \to$ \cc~ will become increasingly important
while for the threshold region, $r^2 \to 1$, the $\epem \to \tso \to$ \cc~ becomes more dominant. 

At the 10.6 GeV c.m. energy of the data \cite{belleddstar}, $r^2 = 0.28$, and ref \cite{braaten} 
finds for the one-gluon exchange (OgE) contribution to the cross sections $\sigma(\psi \chi_0:\psi\chi_1:\psi\chi_2) 
\sim 12:2:3$, 
which contradicts
eq.\rf{smallscalar} based upon factorization and assumption of a $^3\uS_1$ initial state.
In the threshold limit $r^2 \to 1$ the analysis simplifies and comparison between the predictions 
of gluon exchange
and factorization becomes sharpest.
In this limit the $VT$ amplitudes for transversely polarised initial state of
ref.\cite{braaten} satisfy
\be
a[V(-)T(++)]:a[V(0)T(+)]:a[V(+)T(0)] = 1:1/\sqrt{2}:1/\sqrt{6}
\label{vtratios}\ee
in accord with S-wave dominance and the results of Table \ref{helmult}. The relative 
cross-sections from the OgE mechanism
for $\epem \to \psi \chi_{0,1,2}$ in 
vicinity of threshold $r^2 \to 1$ in ref \cite{braaten} become

\be
\sigma(\psi \chi_0:\psi\chi_1:\psi\chi_2) \sim 24:2:3.
\label{oge}
\ee
Compared to the results at higher energy, $r^2 = 0.28$,
the relative sizes of $\psi \chi_1:\psi\chi_2$
have not changed much but there is a significant relative enhancement of $\sigma(\psi \chi_0)$
near threshold.

This prediction, that the cross-section for $\psi \chi_0$ dominates, contrasts with the results 
of factorization near 
threshold. 
For $\tso$ initial state in the $S$-wave
region near threshold

\bea
\label{ratioss}
\sigma(\psi \chi_2) &\to& 0\\ \nonumber
\sigma(\psi \chi_0) &=& \frac{3}{4}\sigma(\psi \chi_1)
\eea
\noindent 
Analogously, for a $\tdo$ initial state 

\bea
\label{ratiodd}
\sigma(\psi \chi_0)& \to &0\\ \nonumber
\sigma(\psi \chi_1) &=& \frac{5}{3}\sigma(\psi \chi_2)
\eea
\noindent which is also utterly unlike the OgE predictions. Finally one may allow for a coherent 
mixture of
$\tso$ and $\tdo$ initial state. Results become model dependent but $\sigma(\psi \chi_0)$ cannot 
be made
larger than both  $\sigma(\psi \chi_1)$ and $\sigma(\psi \chi_2)$. Thus in the region of 
threshold
there appear to be marked differences in the expectations of factorization, eqs 
(\ref{ratioss}),(\ref{ratiodd}) and OgE
eq.(\ref{oge}).

As one increases energy above threshold, for $\tso$ initial state, the $D$-wave decay enables 
$\sigma(\psi \chi_2)$ to 
turn on but
with the amplitude $a[V(-)T(++)] = 0$ or at least small compared to $a[V(0)T(+)]$ and 
$a[V(+)T(0)]$.
This also contrasts with the predictions from OgE where the $[V(-)T(++)]$ amplitude
is the largest for $VT$ production, eq(\ref{vtratios}). A possible contamination comes from 
$\epem \to \tdo \to$ \cc~ 
contributions
which may not be negligible at 6-7GeV c.o.m energies. 
The S-wave decay amplitudes from initial
$\tso, \tdo$ and also from hybrid vector mesons are compared in Table \ref{charmonium}. 
Above threshold where $D$-wave decays are important and $\tso$-$\tdo$ mixing is allowed, results 
are highly
model dependent. While it may be possible to force $\sigma(\psi \chi_0)$ to dominate by suitable 
choice of
mixing angle, this is not expected to hold true as a function of $q^2$. 

Thus if dominance of $\psi \chi_0$
is confirmed over a range of $q^2$ away from threshold, this would support OgE as the dominant 
decay mechanism.
Conversely, if data near threshold confirm $a[V(-)T(++)] \to 0$, this would signal factorization
being dominant. In any event, we anticipate that the relative populations and helicity structures
of $\psi \chi_j$ will vary with $q^2$. We recommend that this be 
investigated in $\epem$ annihilation at super-B factories by means of
ISR to access a range of energies. In particular experiment should 
attempt to measure the spin dependence of $\epem \to \psi \chi_2$ as a function of $q^2$ and 
compare with the analogous 
amplitudes in $e^+e^- \to \omega f_2$.

%

%

\vskip 0.1in

\section{Conclusion}

The factorization property of strong decay triggered by \qq~ creation in spin-triplet, as 
revealed by
lattice QCD, merits further testing. This general feature leads to relations among amplitudes,
which can be used as further tests of this dynamics and to determine the nature of participating 
mesons.
Thus we have identified the following tests.

\begin{enumerate}
\item $\psi(n^3\uS_1)$ decays or $\epem$ annihilation in the 4-5 GeV energy range will not produce
$D^*D_2$ with the tensor meson in helicity two. This tests whether the dynamics
revealed by lattice QCD for light mesons applies more generally for the strong creation of light 
flavours.

\item If confirmed, then the production $\epem \to D^{(*)}D_1$ may be used to determine the axial 
meson mixing
angles in the $^3\uP_1$-$^1\uP_1$ bases.

\item If the mixing angles are known from elsewhere, the pattern of charm pair production can 
identify the
nature of the decaying $\psi$ state. This has an application of immediate relevance in 
determining the
nature of the enigmatic charmonium-like 
structures $Y(4260)$ and $Y(4325)$ and also of $\psi(4415)$. Determining whether the
\cc~ content of these states is $S=0$ (as for a hybrid) or $S=1$ then follows from the relative
production rates of various combinations of charmed mesons, in particular of their $DD_1$ 
branching ratios.

\item The application to light flavours in $\epem$ is less solid, but measurement of the $\omega 
f_2$ amplitudes
as a function of $q^2$ may isolate $\tdo$ resonances in the $\epem$ channel.

\item For the creation of heavy flavours, as in  $\epem \to \psi \chi_j$, empirical and 
theoretical arguments suggest that
production is dominated by a single hard gluon rather than the factorization mechanism. The 
apparent excess of
$\psi \chi_0$ and absence of $\psi \chi_2$ needs establishing. We expect that the pattern of 
$\chi_j$ states
and their helicity amplitudes will vary significantly with $q^2$. We identify the threshold 
region $\epem \to \psi \chi_j$
between 6.5 and 7.5 GeV as particularly promising for determining the relative importance of 
single hard gluon
and strong factorization for heavy flavours.
\end{enumerate}

\vskip 0.5in

We are grateful to E.Swanson for discussions. This work is supported,
in part, by grants from
the Science and Technology Facilities Council, the Oxford University
Clarendon Fund and the
EU-TMR program ``Eurodice'', HPRN-CT-2002-00311.


\end{document}